\documentclass{emulateapj}
\newcommand{\bq}{\begin{equation}}
\newcommand{\eq}{\end{equation}}
\newcommand{\bqa}{\begin{eqnarray}}
\newcommand{\eqa}{\end{eqnarray}}

\newcommand{\Mpc}{\mbox{ Mpc}}
\newcommand{\Mpcden}{\mbox{ Mpc}^{-3}}

\newcommand{\Mpcinv}{\mbox{ Mpc$^{-1}$}}

\newcommand{\kel}{\mbox{ K}}
\newcommand{\mkel}{\mbox{ mK}}

\newcommand{\ergsechz}{\mbox{ erg s$^{-1}$ Hz$^{-1}$}}

\newcommand{\hr}{\mbox{ hr}}

\newcommand{\MHz}{\mbox{ MHz}}

\newcommand{\Msun}{\mbox{ M$_\odot$}}

\newcommand{\hunits}{\mbox{ km s$^{-1}$ Mpc$^{-1}$}}

\newcommand{\bxhi}{\bar{x}_{\rm HI}}
\newcommand{\xhi}{x_{\rm HI}}

\newcommand{\bdtb}{\bar{\delta T}_b}

\newcommand{\hone}{HI }
\newcommand{\htwo}{HII }
\newcommand{\mmin}{m_{\rm min}}
\newcommand{\fcoll}{f_{\rm coll}}
\newcommand{\fesc}{f_{\rm esc}}
\newcommand{\lya}{Ly$\alpha$ }
\newcommand{\lyans}{Ly$\alpha$} % Use this version if Ly\alpha is
\newcommand{\bk}{{\bf k}}

\newcommand{\deriv}{{\rm d}}
\def\VEV#1{\left\langle #1\right\rangle} % This is \VEV{x} => <x>

\begin{document}

\title{The Cross-Correlation of High-Redshift 21 cm and Galaxy Surveys}

\author{Steven R.  Furlanetto\altaffilmark{1} \& Adam Lidz\altaffilmark{2}}

\altaffiltext{1} {Yale Center for Astronomy and Astrophysics, Yale University, 260 Whitney Avenue, New Haven, CT 06520-8121; steven.furlanetto@yale.edu}

\altaffiltext{2}{Harvard-Smithsonian Center for Astrophysics, 60 Garden Street, Cambridge, MA 02138}

\begin{abstract}
We study the detectability of the cross-correlation between 21 cm emission from the intergalactic medium and the galaxy distribution during (and before) reionization.  We show that first-generation 21 cm experiments, such as the Mileura Widefield Array (MWA), can measure the cross-correlation to a precision of several percent on scales $k \sim 0.1 \Mpcinv$ if combined with a deep galaxy survey detecting all galaxies with $m>10^{10} \Msun$ over the entire $\sim 800$ square degree field of view of the MWA.  LOFAR can attain even better limits with galaxy surveys covering its $\sim 50$ square degree field of view.  The errors on the cross-power spectrum scale with the square root of the overlap volume, so even reasonably modest surveys of several square degrees should yield a positive detection with either instrument.  In addition to the obvious scientific value, the cross-correlation has four key advantages over the 21 cm signal alone:  (1) its signal-to-noise exceeds that of the 21 cm power spectrum by a factor of several, allowing it to probe smaller spatial scales and perhaps to detect inhomogeneous reionization more efficiently; (2) it allows a cleaner division of the redshift-space distortions (although only if the galaxy redshifts are known precisely); (3) by correlating with the high-redshift galaxy population, the cosmological nature of the 21 cm fluctuations can be determined unambiguously; and (4) the required level of foreground cleaning for the 21 cm signal is vastly reduced. 
\end{abstract}
  

\section{Introduction}
\label{intro}

The epoch of reionization is one of the landmark events in structure formation, because it defines the moment at which the small fraction of material bound inside galaxies affected each and every baryon in the Universe.  As such reionization is an excellent tracer of the early generations of structure formation, offering a window into the properties of the first stars, the formation of the first quasars, and the growth of galactic systems.  It is also a crucial event because of its effects on the galaxies themselves, suppressing the formation of small systems and affecting the formation of galaxies like our own.  All of this has made the epoch of reionization one of the frontiers of modern astrophysics.

Observations are now beginning to probe reionization, but it remains mysterious.  High-redshift quasars selected from the \emph{Sloan Digital Sky Survey} may indicate a rapidly increasing neutral fraction at $z \sim 6$ \citep{fan06}, although that conclusion is model-dependent \citep{songaila04, lidz06, becker06-tau}.  Specific features in some of the spectra may also point toward a neutral fraction $\xhi \ga 0.1$ along at least some lines of sight \citep{mesinger04, wyithe04-prox, wyithe05-prox, mesinger06-prox}.  On the other hand, the abundance of \lyans-emitting galaxies at $z=6.56$ argues strongly against a predominantly neutral Universe at that time \citep{malhotra04, furl06-lyagal, malhotra06}, and the cosmic microwave background (CMB) polarization observed by the \emph{Wilkinson Microwave Anisotropy Probe} (\emph{WMAP}) \citep{page06} points toward reionization beginning as early as $z \ga 8$--10.  

Disentangling this knot will require new techniques to observe the high-redshift Universe.  Two of the most promising are galaxy surveys and the 21 cm line.  The merits of the former are obvious, because such observations reveal the detailed properties of the ionizing sources.  The 21 cm line is perhaps the  most exciting possibility for directly studying the reionization of the intergalactic medium (IGM); see \citet{furl06-review} for a recent review.  While it remains neutral, the IGM will be a net emitter or absorber of redshifted 21 cm photons from the CMB provided that its spin temperature differs from the CMB temperature, a condition that should be satisfied well before reionization is complete \citep{sethi05, furl06-glob}.  In that case, fluctuations in the 21 cm brightness trace fluctuations in the density, ionized fraction, and spin temperature of the IGM, allowing (in an ideal world) a tomographic reconstruction of the history of structure formation \citep{madau97}.  Unfortunately, the experimental challenges are formidable indeed, especially given the extremely bright Galactic and extragalactic foregrounds at the low frequencies ($\nu < 200 \MHz$) relevant for these observations.  Cleaning these foregrounds will require great care in calibration and sophisticated data analysis algorithms (see the discussion in \S 9 of \citealt{furl06-review}).

One interesting question is how these two datasets can be combined to reveal even more information about the high-redshift Universe. The potential synergy is obvious, given the complementarity between studying the ionizing sources and (to-be-ionized) IGM.  \citet{wyithe06-cross} made a first step in this direction by showing that first-generation 21 cm surveys, together with \emph{existing} galaxy surveys, may be able to distinguish ``inside-out" and ``outside-in" reionization scenarios (in which, respectively, over- and underdense gas is ionized first).  But of course even more is possible.  For example, the ionizing efficiency could vary with galaxy mass (or any other parameter), so that some subset of the galaxy population is responsible for most of reionization.  The structure of the \htwo regions will depend upon such factors \citep{furl05-charsize, mcquinn06}, and we can most efficiently learn about them by comparing the galaxies to the 21 cm pattern.  Furthermore, the small-scale fluctuations in the 21 cm signal depend on the interactions between the galaxies and the cosmic web surrounding them (e.g., \citealt{zahn06-comp,lidz06-ng}).  Cross-correlation also has some useful properties on the data analysis side:  because only a small fraction of the 21 cm foreground originates from high redshifts, cross-correlation with the galaxies (known to be at high-redshift) greatly eases the foreground removal requirements and offers unambiguous confirmation of the cosmological signal.

In this paper, we will study the detectability of this cross-correlation and quantify how well it can be measured with a variety of 21 cm and galaxy surveys.   Note that we will \emph{not} explore the astrophysics content of the cross-correlation:  we will take the simplest possible signal and consider how well one can measure it in realistic experiments.  We defer a detailed examination of the science return to future work (also see \citealt{wyithe06-cross}).  In \S \ref{method}, we describe our simple model for the signal and how we calculate the errors in the cross-correlation.  We present our results for the sensitivity in \S \ref{results}, including the spherically-averaged power, the effect of redshift errors in the galaxy survey, and the sensitivity to redshift-space distortions.  We then consider the effect of 21 cm foregrounds in \S \ref{foreground}, and finally we conclude in \S \ref{disc}.

In our numerical calculations, we assume a cosmology with $\Omega_m=0.26$, $\Omega_\Lambda=0.74$, $\Omega_b=0.044$, $H=100 h \hunits$ (with $h=0.74$), $n=0.95$, and $\sigma_8=0.8$, consistent with the most recent measurements \citep{spergel06}.\footnote{Note that we have increased $\sigma_8$ above the best fit \emph{WMAP} value in order to improve agreement with weak lensing measurements.}  Unless otherwise specified, we use comoving units for all distances.

\section{Method} \label{method}

\subsection{The Signal}
\label{signal}

The 21 cm power spectrum and the galaxy power spectrum are themselves complicated beasts, and we expect their cross-correlation to be even more complex because galaxies source most of the fluctuations in the brightness temperature (especially \htwo regions, whose strong contrast and complex shapes dominate the fluctuations throughout reionization).  Of course, this makes it a magnificent probe of the interactions between galaxies and the IGM; unfortunately, it also makes the signal difficult to model robustly and transparently.  Because we are not concerned here with the detailed astrophysics underlying the cross-correlation but rather in its overall detectability, we will take the simplest possible model for it.  We assume that both the 21 cm brightness and the galaxy distribution trace the linear density power spectrum $P_{\delta \delta}$.  We thus neglect perturbations to the 21 cm signal from ionized regions (as well as inhomogeneous heating and spin temperature coupling).  This is obviously a drastic oversimplification, but it allows us to examine the prospects for detection in a straightforward manner.  We also neglect nonlinear corrections.  For the density field, these become important at $k \ga 5 \Mpcinv$, near the upper range of the scales we will consider.  However, galaxies are so highly biased that nonlinear effects set in even on large scales and may affect the detailed shape of the power spectrum through much of the range we consider.  With these simplifications, the cross-power spectrum between the galaxy field and the 21 cm brightness temperature can be written
\bq
P_{21,g}(k,\mu,z) = (1 + \beta \mu^2)(1 + \beta \mu^2/\bar{b}) \bar{b} \bdtb P_{\delta \delta}(k,z),
\label{eq:sig}
\eq
where $\bk$ is the wavenumber, $\mu$ is the cosine of the angle between $\bk$ and the line of sight, $k=|\bk|$,  $\beta \approx \Omega_m(z)^{0.6}$, $\bar{b}$ is the mean galaxy bias within the sample, $\bdtb \sim 20 \bxhi \mkel$ is the mean 21 cm brightness temperature of the IGM (see \citealt{furl06-review} for details), and $\bxhi$ is the globally-averaged neutral fraction (which we assume to be unity throughout).  $\beta$ accounts for the growth of velocity perturbations relative to those of density; the two factors describe these redshift-space distortions in the 21 cm and galaxy signals, respectively \citep{kaiser87, bharadwaj04-vel, barkana05-vel}.  Note that we will quote results in terms of the power per logarithmic interval, $\Delta^2 = k^3 P(k)/(2 \pi^2)$.

Again, we emphasize that equation~(\ref{eq:sig}) is a naive simplification.  During reionization, galaxies seed \htwo regions, which introduce substantial -- and often dominant -- fluctuations into the 21 cm signal \citep{furl04-bub}.  Reionization will modify our signal in two significant ways.  First, of course, the galaxy positions and the 21 cm brightness will actually be \emph{anti}-correlated because galaxies must sit inside ionized bubbles.  Second, the \htwo regions grow to rather large sizes (easily $\ga 10 \Mpc$), amplifying the signal on relatively large scales (see \citealt{furl04-bub, zahn06-comp, iliev05-sim}).  We will examine these physical effects in future work; for now, our forecasts can be viewed as the ability to rule out the ``null hypothesis" that galaxies and the 21 cm signal both simply trace the underlying density field, or in other words to detect the ionized zones around the sampled galaxy population.  Our fractional errors are actually conservative so long as the \htwo regions do indeed amplify the large-scale signal.

\subsection{Error Estimates}
\label{errors}

Neglecting systematic effects such as foreground subtraction, the errors on a measurement of the 21 cm power (with true value $P_{21}$) at a particular mode $(k,\mu)$ are \citep{mcquinn05-param}
\bq
\delta P_{21}(k,\mu) = P_{21}(k,\mu) + \frac{T_{\rm sys}^2}{B t_{\rm int}}\, \frac{D^2 \, \Delta D}{n(k_\perp)} \, \left( \frac{\lambda^2}{A_e} \right)^2.
\label{eq:dp21}
\eq
Here $T_{\rm sys}$ is the system temperature of the telescope, $B$ is the total bandwidth of the measurement, $t_{\rm int}$ is the total integration time, $\lambda$ is the wavelength of the observation, and $A_e$ is the effective area of each telescope; the last factor is of order unity for antennae optimized to observe at the appropriate redshift.  The distance to the survey volume is $D$ and its radial width is $\Delta D$.  The array geometry enters through the factor $n(k_\perp)$, which is the density of baselines observing at the appropriate transverse wavevector, $k_\perp = (1-\mu^2)^{1/2}k$, normalized so that its integral over the half-plane is the total number of baselines, $N_a(N_a-1)/2$, where $N_a$ is the number of independent elements in the array.  This spans a range defined by the minimum and maximum baselines in the array; for example, $k_{\perp,{\rm max}} = 2 \pi L_{\rm max}/(D \lambda)$, where $L_{\rm max}$ is the maximum baseline distance in the array.  In equation~(\ref{eq:dp21}), the first term represents cosmic variance and the second thermal noise.

The analogous errors on a galaxy survey (with true power spectrum $P_{\rm gal}$) are \citep{feldman94, tegmark97}
\bq
\delta P_{\rm gal}(k,\mu) = P_{\rm gal}(k,\mu) + n_{\rm gal}^{-1} e^{k_\parallel^2 \sigma_r^2},
\label{eq:dpgal}
\eq
where $n_{\rm gal}$ is the mean number density of galaxies in the survey, $k_\parallel=\mu k$ is the component of $k$ along the line of sight, $\sigma_r = c \sigma_z/H(z)$, and $\sigma_z$ is the typical error in each redshift measurement.  Here the first term is again cosmic variance, and the second term is a combination of shot noise ($1/n_{\rm gal}$) and redshift errors, which smear the observed radial fluctuations but leave the transverse power unaffected (see, e.g., \citealt{seo03}).  Given the difficulty of high-redshift galaxy surveys, we allow for the possibility of large photometric redshift errors below.

The effective number density depends on the characteristics of the survey and on the (unknown) galaxy population at high redshifts.  Rather than attempt to model these in detail, we will take a simplistic approach and assume that all dark matter halos with $m>\mmin$ are detectable by the survey.  We use the \citet{press74} mass function; this probably underestimates the number of large halos and so provides a conservative estimate for the errors \citep{jang01,reed03,iliev05-sim, zahn06-comp}.  Our results can easily be rescaled to other mass functions by choosing $\mmin$ to match our number density; we have $n_{\rm gal}=2.4 \times 10^{-3},\,7.3 \times 10^{-6}$, and $7.2 \times 10^{-7} \Mpcden$ for $\mmin=10^{10},\,10^{11}$, and $2 \times 10^{11} \Msun$ at $z=8$.  The mean bias in equation~(\ref{eq:sig}) is then the mean bias of all the galaxies above this threshold, computed using the standard \citet{mo96} formula.  We find $\bar{b} = 7.5,\,11.8$, and $13.7$ for our three surveys.

The error in the cross-correlation for a particular mode is then 
\bq
2 [\delta P_{21,g}^2(k,\mu)] = P_{21,g}^2(k,\mu) + \delta P_{21}(k,\mu) \delta P_{\rm gal}(k,\mu).
\label{eq:dpcross}
\eq
The factor of two comes from only sampling the upper half-plane, because the power spectrum is the Fourier transform of a real-valued function.  In most of the range we consider, the second term dominates by a large factor.  

To this point, we have considered each mode individually; of course, in practice, we will bin them in both $k$ and $\mu$.  The number of modes available in an annulus of width $(\Delta k, \Delta \mu)$ is
\bq
N_m = 2 \pi k^2 \Delta k \Delta \mu \, \left[ \frac{V_{\rm surv}}{(2 \pi)^3} \right],
\label{eq:nm}
\eq
where the last factor is the Fourier space resolution and the survey volume is $V_{\rm surv}= D^2 \Delta D (\lambda^2/A_e)$.\footnote{Here we implicitly assume that the minimum baseline is equal to the radius of one antenna -- i.e., a filled core of antennae.  Then the maximum spatial scale to which the interferometer is sensitive is $\propto D/\sqrt{A_e}$.}   We can then compute the net error within each bin by simply adding the errors in inverse quadrature.  For example, the errors on the spherically-averaged power spectrum are
\bq
\frac{1}{\delta P_{21,g}^2(k)} = \sum_\mu \Delta \mu \,  \frac{\epsilon k^3 V_{\rm surv}}{4 \pi^2} \, \frac{1}{\delta P_{21,g}^2(k,\mu)},
\label{eq:dpsphere}
\eq
where $\epsilon = \Delta k/k$.

The sum over $\mu$ extends over all detectable modes that fit inside the survey volume.  Because we allow modes in only the half-plane, we must have $0 \le \mu \le 1$.  A typical 21 cm survey is (in the flat-sky approximation) a thin rectangular slice of depth $\Delta D \ll D$:  this is much different from a typical galaxy survey, in which the depth is large.  For this reason, not all modes are available for extraction.  In particular, $k_{\parallel,{\rm min}}=2 \pi/\Delta D$, so $\mu_{\rm max}={\rm min}(1,k/k_{\parallel,{\rm min}})$.  In the transverse direction, galaxy surveys are in principle sensitive to infinitely large $k_\perp$, but 21 cm surveys are limited by the maximum angular resolution of the telescope -- determined by $L_{\rm max}$.  Thus $\mu_{\rm min}^2={\rm max}(0,1-k^2_{\perp,{\rm max}}/k^2)$.

\section{Results}
\label{results}

We are now in a position to estimate errors on the cross-power spectrum of the 21 cm sky and the galaxy field.  We will use a fiducial 21 cm survey at $z=8$ with similar parameters to the Mileura Widefield Array (MWA) Low Frequency Demonstrator (as in \citealt{mcquinn05-param,bowman05-param}).  We take a total area $A_{\rm tot}=7000~{\rm m}^2$ spread over $N_a=500$ antennae, distributed in a circle of radius $0.75~{\rm km}$.  We assume that each element is $4~{\rm m}$ wide, and that they are closely packed within a filled core and distributed like $r^{-2}$ outside of that core.  We take $t_{\rm int}=1000 \hr$, $B=12 \MHz$ (note that this may be large enough for evolutionary effects to become important across the band, though we ignore that possibility here; \citealt{mcquinn05-param, barkana05-cone}), and $T_{\rm sys}=440 \kel$.  We divide the measurement into bins of logarithmic width $\epsilon=0.5$.  We will also show some estimates for a fictional array with a square kilometer of collecting area; although we will refer to it as the Square Kilometer Array (SKA), our choices do not actually correspond to any specific design proposal.  We take $A_{\rm tot}=1~{\rm km}^2$ spread over $N_a=5000$ antennae, with a maximum baseline of $2.5 \, {\rm km}$.  The antennae are distributed in a similar pattern to those of the MWA.

We will also assume that the accompanying galaxy survey covers the entire volume of the 21 cm survey.  This may be difficult in practice, because the field of view of the MWA is $\sim 800$ square degrees at this wavelength.  Fortunately, using equation~(\ref{eq:dpsphere}), it is relatively easy to transform our results to a more realistic case where the galaxy survey subtends only a fraction of the 21 cm survey.  By decreasing the effective survey volume, this increases the error according to $\delta P_{21,g} \propto V_{\rm surv}^{-1/2}$.  Thus confining the galaxy survey to a small but contiguous field simply increases the errors by the corresponding factor, and of course it also prevents the cross-correlation from measuring any modes wider than the galaxy field.  However, the latter effect is not nearly as important as one might expect, because (as we will see below) modes wider than the redshift depth are lost anyway.  

A second strategy is sparse sampling:  to distribute the galaxy observations across the entire 21 cm field, but with a small filling factor.  In this case, the $V_{\rm surv}^{-1/2}$ scaling still approximately holds, although the window function complicates the constraints.  In general, the cost of sparse sampling is that power can be aliased from high-$k$ modes.  The optimal survey design would depend on the details of the power spectrum (see, e.g., \citealt{heavens97, kaiser98} for discussions of similar issues in galaxy redshift surveys and weak lensing).

\subsection{Spherically-Averaged Cross-Power Spectra}
\label{sphere}

\begin{figure}
\plotone{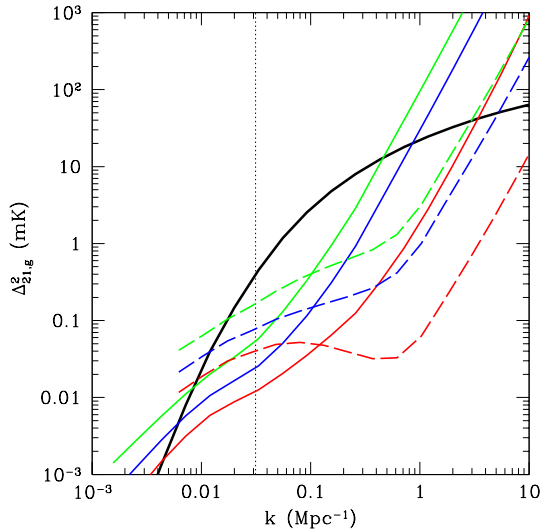}
\caption{Expected errors on the cross-power spectrum.  The solid black curve is $\Delta^2_{21,g}$ for our simple model in which both galaxies and density fluctuations trace the underlying dark matter power spectrum.  The three thin solid curves show the errors for the MWA combined with a galaxy survey reaching $\mmin=10^{10},\,10^{11}$, and $2 \times 10^{11} \Msun$ (from bottom to top); the dashed curves show the corresponding errors for our SKA.  In each case, we assume 1000 hr observations at $z=8$.  The vertical dotted line shows the foreground cut for this 21 cm survey (corresponding to $B=12 \MHz$); modes leftward of this line are removed from the 21 cm power spectrum during foreground cleaning.}
\label{fig:cross}
\end{figure}

Figure~\ref{fig:cross} shows our estimate for the spherically-averaged cross-correlation signal, in the context of the simple model of \S \ref{signal}.\footnote{Note that the signal shown here assumes $\mmin=10^{10} \Msun$; it is nearly proportional to the mean bias of the galaxies and so actually depends on the particular galaxy survey (see \S \ref{errors}).}  In all cases we assume perfect redshift information for the galaxy survey.  In our simple model, the signal is proportional to the linear power spectrum, amplified by the galaxy bias and redshift space distortions.  The thin solid curves show our error estimates for the MWA, while the dashed curves show them for the SKA.  Within each set, the curves assume that the associated galaxy survey has $\mmin=2 \times 10^{11},\,10^{11},$ and $10^{10} \Msun$, from top to bottom. These have $\sim 2400,\,24,000,$ and 8 million galaxies over the survey volume, respectively.

%%%%%%%%%%FIGURE 1:  Errors as a function of survey depth, absolute terms

Ignoring systematics, the MWA would provide a measurement over the range $0.005 \Mpcinv \la k \la 1 \Mpcinv$ (in about ten independent bins).  The SKA could reach even smaller scales -- although it is limited on larger scales because our version actually has a somewhat smaller field of view than the MWA.  However, as with the 21 cm power spectrum, this range is misleading because of foregrounds \citep{mcquinn05-param}.   The dotted curve shows the wavenumber corresponding to $k_{\parallel,{\rm min}}$.  On scales $k<k_{\parallel,{\rm min}}$, one must take into account discreteness in the Fourier transform -- in particular, the only modes with $k_\parallel < k_{\parallel,{\rm min}}$ that are permitted in our thin rectangular slice actually have $k_\parallel=0$ and so correspond to modes along the plane of the sky.  Such modes cannot be separated from fluctuations in the astrophysical foregrounds; thus, in reality, only modes rightward of the dotted line are relevant.  Even with the relatively large bandwidth we have chosen here, this substantially reduces the accessible range of scales.

\begin{figure*}
\plottwo{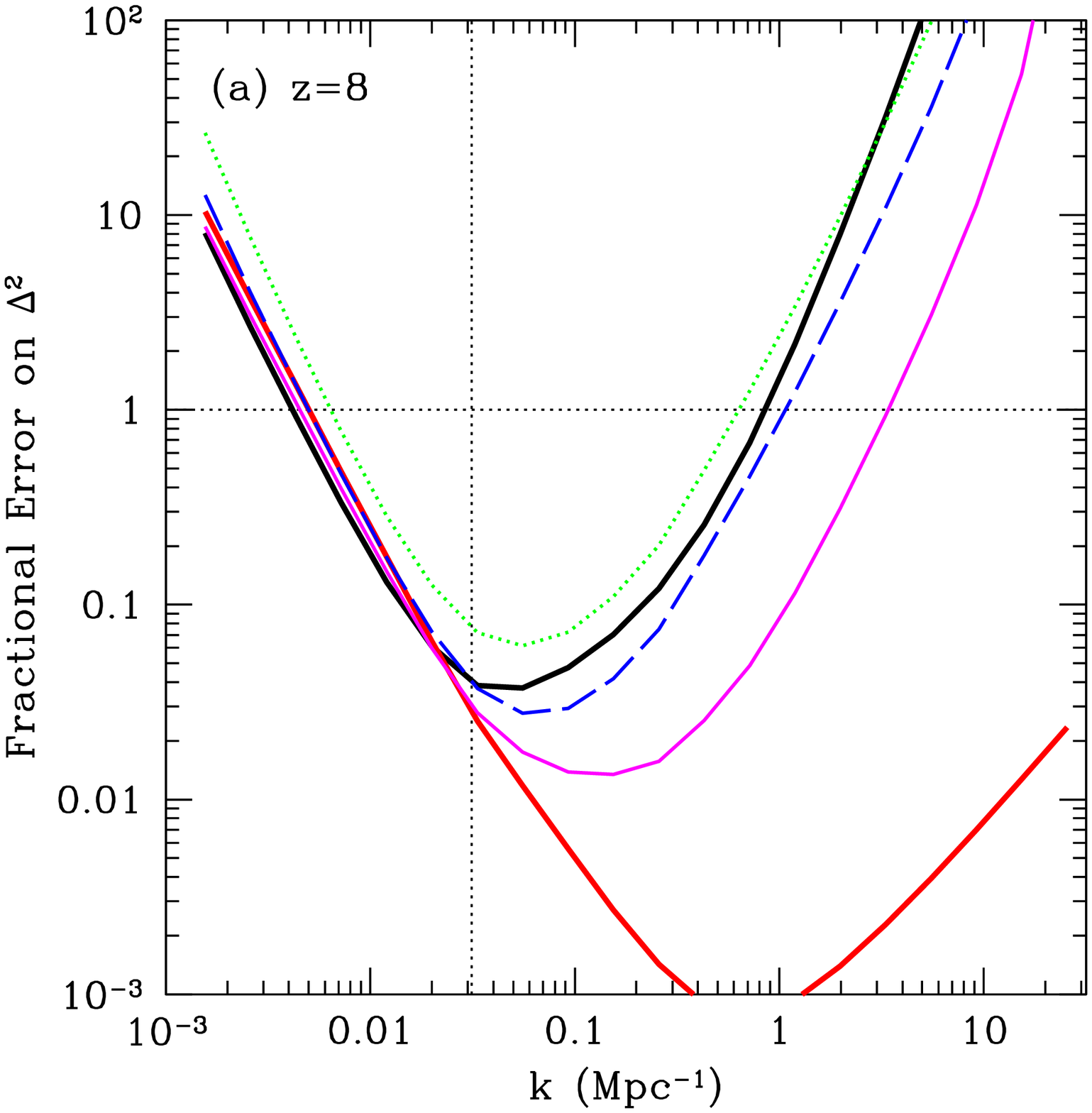}{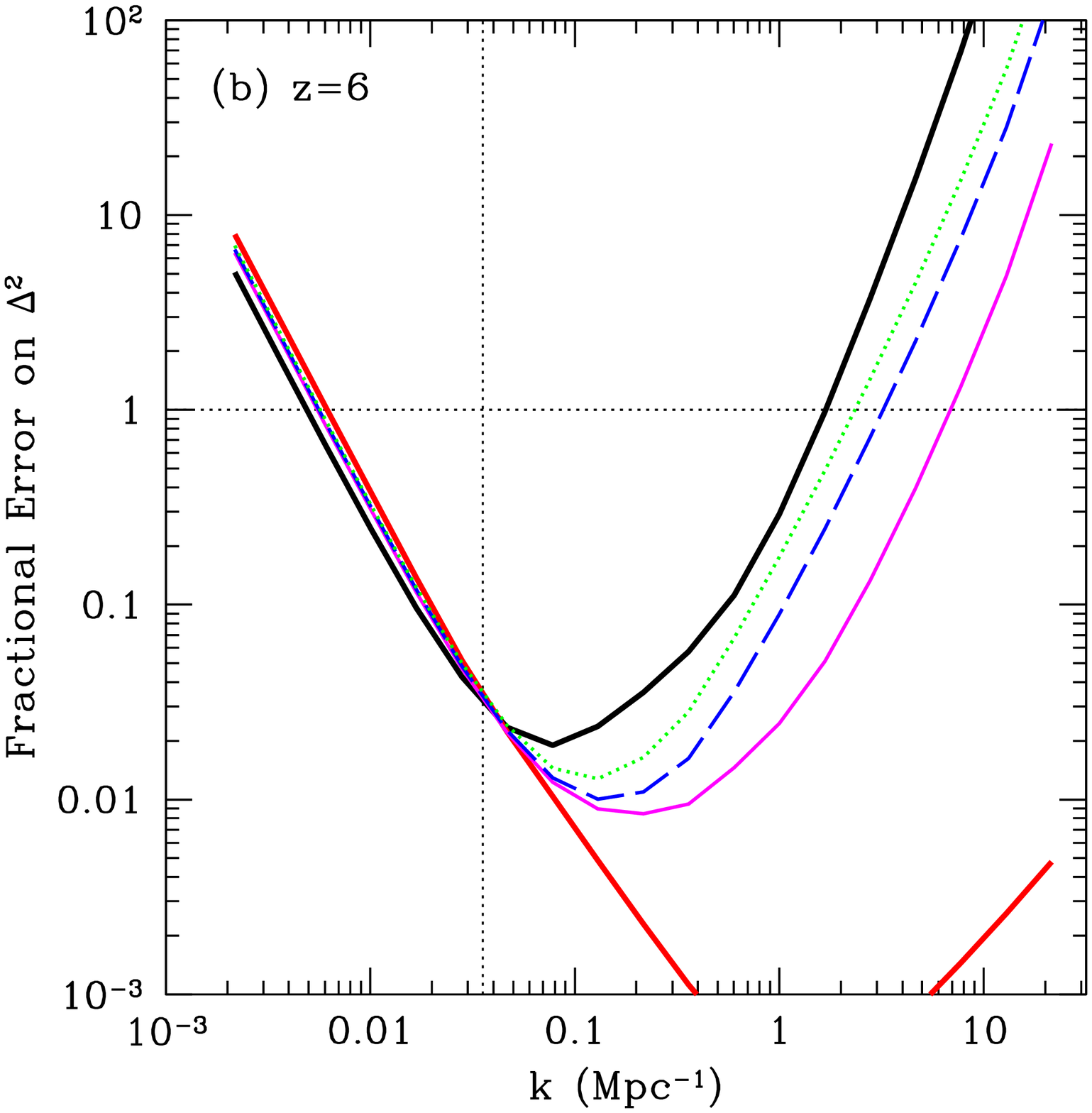}
\caption{Fractional errors on power spectra for the MWA at \emph{(a)} $z=8$ and \emph{(b)} $z=6$.  The upper thick solid curves are for 21 cm surveys with our fiducial MWA parameters, while the lower thick solid curves are for galaxy surveys with $\mmin=10^{10} \Msun$.  The thin curves are for the cross-correlation between the two, with $\mmin=10^{10},\,10^{11}$, and $2 \times 10^{11} \Msun$ (solid, dashed, and dotted curves, respectively).  The vertical dotted line shows the foreground cut for this 21 cm survey (corresponding to $B=12 \MHz$).}
\label{fig:cross-rel}
\end{figure*}

Nevertheless, the cross-correlation is still a promising probe.  Figure~\ref{fig:cross-rel}\emph{a} shows the error for the MWA in fractional terms; the thin solid, dashed, and dotted curves are identical to the three MWA-based surveys shown in Figure~\ref{fig:cross}.  Accuracy near one percent can be achieved when the MWA is combined with a deep and wide galaxy survey.  This is sufficiently precise that the galaxy survey need not span the entire field of view.  For example, reducing the areal coverage to $\sim 1$ square degree would still permit a marginal detection of the signal (especially if it is amplified by large ionized regions).  This is already achievable at slightly lower redshifts (see below) and would offer invaluable information about the high-redshift Universe, as well as a useful confirmation of the 21 cm signal's cosmological origin.  Moreover, even this relatively small size is perfectly adequate for recovering the full range of available $k$-modes.  A one square degree survey would have $k_{\perp,{\rm min}} \sim 0.04 [10/(1+z)]^{0.2} \Mpcinv$, rather near the inevitable cutoff from foregrounds anyway.

%%%%%%%%%%FIGURE 2:  Relative errors 

If the 21 cm data is to be combined with a smaller galaxy survey, the extra field of view is essentially wasted.  This can help to drive the design philosophy of 21 cm arrays:  clearly for this purpose it is better to go deep over a small area rather than to simply add field of view.  For example, an instrument like LOFAR is well-matched in this regard.  \citet{mcquinn05-param} showed that the sensitivities of the MWA and LOFAR to $P_{21}$ are nearly identical, although the field of view of LOFAR is only $\sim 50$ square degrees (possibly split into several independent beams).  It makes up the difference in sampled volume with its larger collecting area (about an order of magnitude larger than the MWA), so that each field is measured much more precisely.  Thus, for a small field galaxy survey, LOFAR would have errors $\sim \sqrt{50/800} \sim 0.25$ smaller than the MWA.  (LOFAR also has larger baselines than the MWA, allowing it to probe deeper in $k$ space.)

The shapes of these error curves can be understood through comparison to the thick curves, which show the fractional errors on the associated measurement of the 21 cm power spectrum (upper thick curve) and the galaxy power spectrum (lower thick curve), again assuming $\mmin=10^{10} \Msun$ and $\sigma_z=0$.  The large-scale errors are nearly identical between the two; this is because the uncertainty is dominated by cosmic variance at small $k$, and the survey volumes are assumed to be identical.  The 21 cm survey reaches peak sensitivity at $k \sim 0.05 \Mpcinv$, near the foreground cutoff.  At smaller scales, the errors increase rapidly for two reasons.  First, the effective surface brightness sensitivity rapidly worsens, because the baseline density decreases.  Second, the longest baseline of the MWA corresponds to $k_{\perp,{\rm max}} \approx 0.6 \Mpcinv$.  For $k > k_{\perp,{\rm max}}$, only modes that are inclined relative to the sky can be measured, so the effective sampling decreases rapidly toward smaller scales \citep{mcquinn05-param, bowman05-param}.  On the other hand, galaxy surveys (at least with $\sigma_z=0$) have infinitely good resolution in every direction, so they remain accurate to much smaller scales (where shot noise takes over).  

From equation~(\ref{eq:dpcross}), we would naively expect the errors on the cross-correlation to be approximately the geometric mean of $\delta P_{21}$ and $\delta P_g$.   While this does appear to be the case near and below the foreground cut, the noise on smaller scales actually increases nearly as rapidly for the cross-correlation as for the 21 cm power spectrum itself.  The reason is the mode sampling:  beyond $k_{\perp,{\rm max}}$, the 21 cm array can only measure a small fraction of the modes, and of course only the measured modes can be correlated with the galaxy information.  Thus, while $P_{21,g}$ can be measured to scales several times smaller than the 21 cm measurement, the improvement is less than may have been hoped and (regardless of the galaxy survey) there is a fixed limit at small scales.  This suggests that increasing the maximum baseline length would offer quite substantial improvements in the cross-correlation measurement on small scales, even if $n(k_\perp)$ is relatively sparse -- as indeed we see in the SKA curves in Figure~\ref{fig:cross}.

Nevertheless, $P_{21,g}$ has two advantages over measurements of $P_{21}$.  First, leveraging the galaxy information increases the signal-to-noise by a factor of a few on scales where cosmic variance can be ignored (at least for first-generation surveys like the MWA; for SKA surveys, the 21 cm errors are already comparable to those in the galaxy survey, so the signal-to-noise is similar).  Second, with a deep survey it extends the range of useful $k$ by a factor of several.  This is important for two reasons.  First, Figure~\ref{fig:cross-rel} shows that the dynamic range (in $k$-space) of the 21 cm measurement is only about an order of magnitude.  Any features from reionization are expected to be relatively broad, so identifying them unambiguously will be relatively difficult \citep{furl04-bub}.  Extending the range by even a modest factor will be useful in interpreting the data.  Second, many of the details of the interactions between the ionizing sources and the IGM -- such as recombinations -- are hidden in the small-scale power \citep{zahn06-comp}.

Figure~\ref{fig:cross-rel}\emph{b} shows the corresponding estimates for $z=6$, just below our current lower limits on the redshift of reionization (and near the upper limits of existing galaxy surveys).  Here we assume $T_{\rm sys}=250 \kel$, offering modest improvements to the 21 cm measurements.  The expected galaxy number density also increases (our surveys now have $\sim 7.2 \times 10^4,\, 3.8 \times 10^5$, and $3 \times 10^7$ galaxies in them).  This presents the most optimistic case for detecting the cross-correlation (by assuming that the 21 cm signal persists to the lowest possible redshift) and can be compared with the estimates of \citet{wyithe06-cross}.  They use a model for the cross-correlation in the presence of ionized regions to show that the ``zero-point" offset between the mean brightness temperature in pixels with and without galaxies is measurable with the MWA and the existing Subaru deep field of \lyans-selected galaxies at $z=6.56$ \citep{shimasaku05, kashikawa06}, spanning $\sim 0.25$ square degrees and $\Delta z=0.11$ (corresponding to $\Delta \nu \approx 2.5 \MHz$).  

We can use our results to estimate the detectability of the cross-correlation with such a ``minimal" galaxy survey.  After correcting for contamination from lower redshift objects, the field contains $\sim 36$ galaxies \citep{kashikawa06}; we would therefore expect $\sim 5 \times 10^5$ galaxies in the entire MWA field of view.  To within the limits of our simple model, the Subaru Deep Field is therefore similar to our $\mmin=10^{11} \Msun$ estimates.  Scaling our results by the ratio of the volumes, we expect a fractional error $\sim 1.3$ in the best-measured bins.  Because \htwo regions will enhance the signal, we therefore expect that even existing surveys can provide a basic detection of the cross-correlation and offer interesting complementary information to the 21 cm survey itself, as in \citet{wyithe06-cross}, provided of course that the IGM is still significantly neutral at $z \sim 6.6$.

\subsection{Redshift Errors}
\label{zerr}

\begin{figure}
\plotone{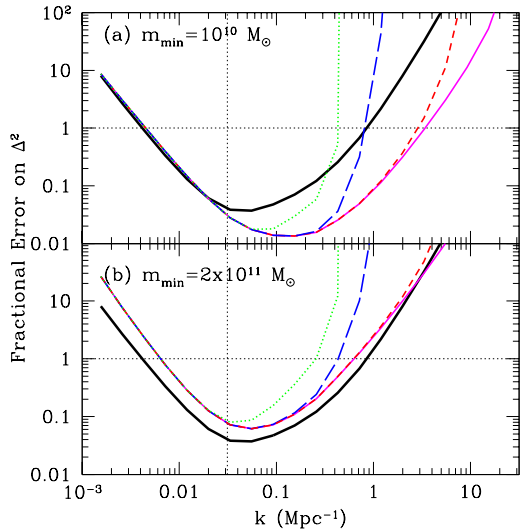}
\caption{Sensitivity to the cross-correlation when redshift errors are included.  In each panel, the solid black curve is for our fiducial 21 cm survey with the MWA at $z=8$.  The others show the errors on the cross-correlation, with the solid, short-dashed, long-dashed, and dotted curves assuming $\sigma_z=0,\,0.001,\,0.01$, and $0.1$, respectively.  We vary $\mmin$ between the two panels as shown.  The vertical dotted lines show the foreground cut for this 21 cm survey (corresponding to $B=12 \MHz$). }
\label{fig:zerr}
\end{figure}

To this point, we have correlated with spectroscopic surveys with infinitely good resolution.  Figure~\ref{fig:zerr} shows the effect of imperfect redshift measurements, again using our fiducial measurements at $z=8$.  The thick solid lines show the errors in the 21 cm power spectrum measurement and are identical between the two panels.  The other curves show errors on $P_{21,g}$ for $\mmin=10^{10} \Msun$ (Fig.~\ref{fig:zerr}\emph{a}) and $\mmin=2 \times 10^{11} \Msun$ (Fig.~\ref{fig:zerr}\emph{b}).  The thin solid, short-dashed, long-dashed, and dotted curves assume $\sigma_z=0,\,0.1\%,\,1\%$, and $10\%$, respectively.  Obviously these errors have no effect on the measurement at large scales, because they only smear out small-scale power.  Naively, one might expect their effects at small scales to be small as well: for the galaxy power spectrum itself, redshift errors have a relatively modest effect, because transverse modes are entirely unaffected.  

%%%%%%%%%%FIGURE 3:  Errors as a function of redshift precision

However, Figure~\ref{fig:zerr} shows that they have a more dramatic effect on cross-correlation measurements.  Again, this is because of the unusual mode sampling in the 21 cm observations.  For wavenumbers larger than that corresponding to $\sigma_z$ (here $k_z \sim 1/\sigma_r \sim 0.3 \Mpcinv$ for $\sigma_z=0.01$) the galaxy survey begins to lose sensitivity to the line of sight modes.  Unfortunately, these line of sight modes are precisely those which the 21 cm survey measures best:  most obviously, a 21 cm telescope is confined to line-of-sight modes on scales above $k_{\perp, {\rm max}}$.   Only when $k < k_{\perp,{\rm max}}$ \emph{and} $k<k_z$ are measurements possible, because in this regime both surveys can use the pristine transverse modes to make measurements.  With any realistic errors from photometric redshifts, $k_z < k_{\perp,{\rm max}}$ -- even for a small telescope like the MWA -- so errors on the small-scale power spectrum will blow up.  Note that this cannot be helped with a deeper galaxy survey:  the exponential cutoff at $k_z$ is much more severe than the slow increase in shot noise (see eq. \ref{eq:dpgal}).  As a result, redshift errors are ``optimally" configured to destroy the cross-correlation measurement.  Accurate spectroscopic redshifts will be necessary to take full advantage of the cross-correlation.

The ultimate limit on redshift accuracy is set by internal motions within the galaxies; if, for example, \lya is used to measure the redshift, winds may displace the line in redshift space by several hundred km s$^{-1}$ (e.g., \citealt{shapley03}).  Such winds would cause $\sigma_z \sim 10^{-3}$, shown by the short-dashed lines in Figure~\ref{fig:zerr}.  Fortunately, these errors are small enough that they do not significantly degrade the  cross-correlation measurement.

\subsection{The Anisotropic Power Spectrum}
\label{angle}

According to equation~(\ref{eq:sig}), redshift-space distortions introduce anisotropy into these power spectra.   In the previous sections we have ignored this anisotropy by averaging over the line of sight angle $\mu$.  We did so  because the sensitivity of first generation 21 cm experiments is heavily weighted along the frequency axis, so they span only a small domain in $\mu$ and will have difficulty extracting the anisotropic components  \citep{mcquinn05-param}.  Cross-correlation with galaxy surveys can improve these constraints by adding sensitivity on small scales, although we still must contend with the hard limit from $k_{\perp,{\rm max}}$.

\begin{figure}
\plotone{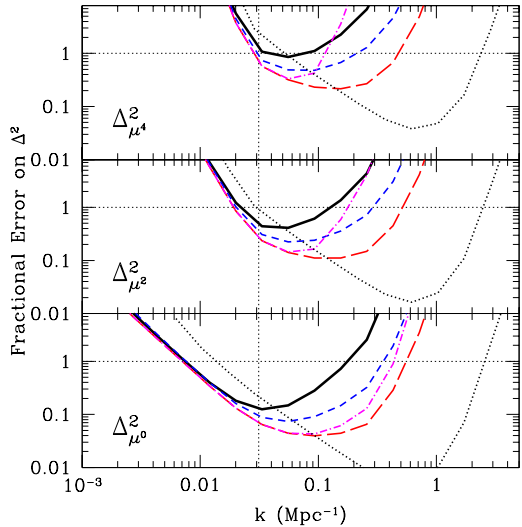}
\caption{Fractional errors on the three $\mu$-dependent components of the power spectrum.  In each panel, the solid black curve is for our fiducial 21 cm survey with the MWA at $z=8$.  The others show the errors on the cross-correlation.  The long-dashed lines take $\mmin=10^{10} \Msun$ and $\sigma_z=0$, the short-dashed lines take $\mmin=10^{11} \Msun$ and $\sigma_z=0$, and the dot-dashed lines take $\mmin=10^{10} \Msun$ and $\sigma_z=0.1$.   The dotted curves show the errors using the SKA, assuming $\mmin=10^{10} \Msun$ and $\sigma_z=0$.  The vertical dotted lines show the foreground cut for this 21 cm survey (corresponding to $B=12 \MHz$).  }
\label{fig:angle}
\end{figure}

To compute the resulting errors on the anisotropic power spectrum, we write
\bq
P_{21,g}(k,\mu) = \mu^0 P_{\mu^0}(k) + \mu^2 P_{\mu^2}(k) + \mu^4 P_{\mu^4}(k)
\label{eq:pmu}
\eq
and use the parameter set $(P_{\mu^0},\,P_{\mu^2},\,P_{\mu^4})$ at each wavenumber $k$ in the Fisher matrix analysis (following \citealt{mcquinn05-param}).  Figure~\ref{fig:angle} shows the resulting fractional errors.  The solid black line is for the MWA 21 cm survey alone (at $z=8$).  After foreground cleaning, it can measure the isotropic component to a precision of $\sim 10\%$ over a limited range of scales, but the $\mu^2$ and $\mu^4$ components will only be weakly constrained (at best).  

%%%%%%%%%%FIGURE 4:  Anisotropic errors

The other curves show the analogous errors on the cross-correlation.  The long-dashed curve assumes a galaxy survey with $\mmin=10^{10} \Msun$ and perfect redshift information.  All of the constraints are markedly better than for $P_{21}$ itself, extending the useable range to smaller scales (by a factor of a few) and also increasing the peak precision by at least a factor of two.  For the MWA, $k_{\perp,{\rm max}} \approx 0.6 \Mpcinv$:  as expected, our sensitivity to the anisotropic components declines rapidly past this point.  The galaxy survey helps to pull out the $\mu$-dependence in the presence of noise, but of course it cannot reach beyond the angular resolution of the 21 cm array where no modes with small $\mu$ are available at all.

The short-dashed curves in Figure~\ref{fig:angle} take a shallower survey with $\mmin=10^{11} \Msun$ and $\sigma_z=0$, while the dot-dashed curves take $\mmin=10^{10} \Msun$ and $\sigma_z=0.1$.  The isotropic component is relatively unaffected in both cases, but redshift errors at this level do significantly degrade the ability to constrain the $\mu^2$ and $\mu^4$ terms.  Note, however, that $1\%$ redshift errors have almost no effect on the measurement, because they distort the spectrum only at $k \ga k_{\perp,{\rm max}}$ anyway.

Finally, the dotted curve shows the errors on the three components for a cross-correlation between a galaxy survey (with $\mmin=10^{10} \Msun$ and $\sigma_z=0$) and an SKA field.  The constraints are worse at small $k$ (because the sampled volume is smaller) but show a dramatic improvement at larger $k$.  The SKA has significantly larger baselines and retains sensitivity all the way to $k \sim 3 \Mpcinv$ in all the angular components.  In this case, the errors on $P_{21,g}$ and $P_{21}$ are actually comparable, so the improvements offered by the cross-correlation are much less significant from a data analysis perspective -- although of course their astrophysical content is still complementary.

\section{Foreground Contamination}
\label{foreground}

To this point we have focused on the improved sensitivity offered by the cross-correlation.  It has the additional advantage of easing the requirements for foreground removal, because the only foregrounds that will survive the cross-correlation are those arising from the cosmological volume of the galaxy survey:  free-free and synchrotron emission from the high-redshift galaxies.  Here we will estimate how strong this contamination is and show that even the simplest foreground removal scheme should adequately remove the residuals.

We will begin by calculating the free-free contamination following the method of \citet{oh03}.  The free-free emissivity can be well-fit by 
\bq
\epsilon_\nu = \epsilon_0 n_e^2 T_4^{-0.35}
\label{eq:enu}
\eq
where $\epsilon_0=3.2 \times 10^{-39}$ ergs cm$^3$ s$^{-1}$ Hz$^{-1}$ and $T_4$ is the electron temperature in units of $10^4 \kel$.  The total free-free luminosity of a single galaxy is the volume integral over all its \htwo regions, $\propto \int n_e^2 \deriv V$.  But of course the total recombination rate in the galaxy is also $\propto \int n_e^2 \deriv V$ (at least ignoring heavy elements).  This allows us to relate the free-free luminosity $L_{\rm ff}$ to the total production rate of ionizing photons if we assume that recombinations are in equilibrium with ionizations:
\bq
\dot{N}_{\rm rec} = \alpha_B \int \deriv V n_e^2 \approx (1-\fesc) \dot{N}_{\rm ion}
\label{eq:nion}
\eq
where $\dot{N}_{\rm rec}$ and $\dot{N}_{\rm ion}$ are the mean rate of recombinations and ionizations in this galaxy, $\alpha_B$ is the case-B recombination coefficient, and $\fesc$ is the escape fraction of ionizing photons.

Thus the mean brightness temperature of free-free emission from our survey volume is simply related to the total ionizing rate.  We assume that the emissivity of ionizing photons is proportional to the total rate at which gas collapses onto galaxies,
\bq
\epsilon_{\rm ion} = f_\star N_{\gamma b} \frac{\bar{\rho}_b}{m_p} (1+z) H(z) \left| \frac{\deriv \fcoll}{\deriv z} \right|,
\label{eq:ionemiss}
\eq
where $\bar{\rho}_b$ is the mean comoving density of baryons, $\fcoll$ is the fraction of matter in galaxy-sized halos, $f_\star$ is the star formation efficiency, and $N_{\gamma b}$ is the number of ionizing photons produced per stellar baryon.  Thus, the mean brightness temperature is (suppressing the temperature dependence of $\epsilon_\nu$ and $\alpha_B$)
\bq
\bar{\delta T}_{\rm ff} \equiv {\mathcal T} \fcoll = \frac{c^2}{2 k_B \nu^2} \, \frac{\epsilon_0}{\alpha_B} \, \frac{(1+z)}{4 \pi d_L^2} \frac{V_{\rm surv}}{\Delta \Omega_{\rm surv}} \, (1-\fesc) \epsilon_{\rm ion},
\label{eq:bdtff}
\eq
where ${\mathcal T}$ is the mean brightness temperature if all the baryons were inside galaxies.  (Note that $\bar{\delta T}_{\rm ff}$ is actually independent of $\Omega_{\rm surv}$, although it does depend on the radial depth.)  For reference, at $z=8$, assuming $f_\star=0.1$, $\fesc=0.1$, and $N_{\gamma b}=4000$ (as appropriate for Population II stars), we find $\bar{\delta T}_{\rm ff} = 4.5 \mkel$ and ${\mathcal T}=161 \mkel$.

Because the free-free spectrum is featureless, it contains no radial information and is most easily described in terms of the angular power spectrum.  In our notation, the two components (clustering and Poisson) are then (e.g., \citealt{peebles80})
\bqa
C_l^{\rm c} & = & {\mathcal T}^2 \fcoll^2 w_l,
\label{eq:clclus} \\
C_l^{P} & = & \frac{\Delta \Omega_{\rm surv}}{V_{\rm surv}} {\mathcal T}^2 \int \deriv m \left( \frac{m}{\bar{\rho}} \right)^2 n(m).
\label{eq:clp}
\eqa
Here $n(m)$ is the halo mass function, the integration extends over all galaxies, and $w_l$ is the appropriate coefficient in the Legendre expansion of the angular correlation function.  We will take $w(\theta) = (\theta/\theta_0)^{-0.8}$ \citep{oh03} where $\theta_0=2'$ is the correlation length.  Then 
\bq
l^2 w_l = 4.85 (l \theta_0)^{0.8}.
\label{eq:wl}
\eq
We will also define $l_0 \equiv 1/\theta_0 = 1719$.  In equation~(\ref{eq:clp}), we have assumed that the free-free luminosity (and hence the ionizing luminosity) of each galaxy is proportional to its mass.  Note that we have not excised any of the bright point sources here; doing so would effectively impose a finite limit on the integrals over halo mass.  Evaluating these for our fiducial survey parameters, we find
\bqa
\left( \frac{l^2 C_l^c}{2 \pi} \right)^{1/2} & \approx & 4.0 \left( \frac{l}{l_0} \right)^{0.8} \mkel,
\label{eq:clc-eval} \\
\left( \frac{l^2 C_l^P}{2 \pi} \right)^{1/2} & \approx & 0.034 \left( \frac{l}{l_0} \right)^{2} \mkel.
\label{eq:clp-eval}
\eqa

The free-free contamination is straightforward to estimate and provides a \emph{minimal} level of contamination, but it is likely to be much smaller than the cumulative synchrotron emission from the same galaxies (produced by fast electrons in their interstellar media).  Unfortunately, this component is also much more difficult to estimate robustly, because it depends on the details of magnetic field generation and cosmic ray acceleration inside of galaxies.  For a simple estimate, we assume that the observed correlation between synchrotron luminosity and star formation rate (SFR) in nearby galaxies applies equally well at high redshifts \citep{yun01}:  
\bq
L_{1.4 {\rm GHz}} = 1.7 \times 10^{28} \left( \frac{{\rm SFR}}{{\rm M}_\odot \, {\rm yr}^{-1}} \right) \ergsechz,
\label{eq:sfr-synch}
\eq
where $L_{1.4 {\rm GHz}}$ is the specific synchrotron luminosity at a (rest) frequency of 1.4 GHz (fortunately, just the frequency we need for the contamination from the survey volume).  The origin of this relation is unclear; one possibility is that the magnetic fields in starbursts are well above their minimal-energy value \citep{thompson06}.  Fortunately, for our simple estimate, its origin is not so important, and we will just use the empirical relation.  In any case, by assuming that the star formation rate tracks the rate at which gas accretes onto galaxies, equation~(\ref{eq:sfr-synch}) allows us to associate the synchrotron volume emissivity to $\fcoll$ just as with free-free emission; we find
\bq
\frac{\bar{\delta T}_{\rm syn}}{\bar{\delta T}_{\rm ff} } \sim \frac{10}{1-\fesc}
\label{eq:synch-ff}
\eq
for our fiducial parameters.  Thus, at least in the simplest models, the synchrotron foreground will be about an order of magnitude larger than the free-free foreground.  Because both originate from the same source population, and because each is proportional to the star formation rate, their fluctuation spectra will also have identical shapes.

The fluctuation amplitude expected from the 21 cm line is a few mK; thus equation~(\ref{eq:clc-eval}) shows that the residual contamination in the \emph{angular} power spectrum is at least comparable to the 21 cm signal, and probably several times larger thanks to the synchrotron component.  If we only had information at one frequency, there would be no way to separate these foregrounds and recover the cosmological signal.\footnote{Actually, because the foreground contamination at a single frequency is proportional to the radial width of the survey, while the 21 cm fluctuation power at that frequency is independent of the depth, the foreground contamination can be minimized by correlating with narrow redshift slices in the galaxy survey.  However, such a method would eliminate modes along the line of sight and so is not as useful.}   But, with multichannel measurements,  we can take advantage of the fact that both synchrotron and free-free emission have smooth, nearly power-law spectra, while the 21 cm background varies rapidly.  Conceptually, we can therefore choose one frequency slice and use it to calibrate the foreground contamination for all frequencies along each direction \citep{zald04,morales04}.  The remaining fluctuation power will then be the sum of the 21 cm signal, errors in beam calibration, and deviations from the fit value to the foregrounds.  These deviations are caused by variations in the spectral indices of sources in different pixels; if all the pixels had the same spectral index, this simple scheme would be perfect.

In reality, the fit is done in Fourier space rather than real space, so we subtract a constant value from each $\bk_\perp$ ``pixel."  In the process, the $k_\parallel=0$ modes are lost (they correspond to modes along the plane of the sky), but those with $k_\parallel>0$ suffer only minor degradation.\footnote{To this point we have ignored discreteness in the Fourier decomposition.  In reality, because the survey volume is a narrow slice, the only modes that \emph{can} exist with $k_\parallel$ smaller than the radial width of the survey have $k_\parallel=0$.  This is why, in Figs.~\ref{fig:cross}--\ref{fig:angle}, we argued that all modes with $k<k_{\parallel,{\rm min}}$ will be lost in the cleaning.}  We can roughly quantify this residual power by estimating the degree of correlation for the foregrounds in two nearby frequency channels $\nu_1$ and $\nu_2$ \citep{zald04}
\bq
I_l(\nu_1, \nu_2) \equiv \frac{C_l(\nu_1,\nu_2)}{\sqrt{ C_l(\nu_1,\nu_1) C_l(\nu_2,\nu_2)}} \approx 1 - \frac{(\delta \zeta)^2}{2} {\rm ln}^2(\nu_1/\nu_2),
\label{eq:il}
\eq
where in the simplest models (in which Poisson fluctuations dominate) $\delta \zeta$ is the scatter in the spectral indices of sources in the beam.  For free-free sources, variations originate from the electron temperature; however, even allowing electron temperatures across the entire range $10^4$--$10^5 \kel$, $\delta \zeta \sim 0.03$ \citep{santos05}.  The dispersion in synchrotron spectra may be much larger, $\delta \zeta \sim 0.2$ \citep{cohen04}, although these measurements spanned 74 MHz--1.4 GHz.  We are only concerned with the variations over a much smaller interval spanning $\sim 100 \MHz$ (in the rest frequency).  Some of the dispersion over the observed frequency range is likely  in the locations or magnitudes of spectral breaks well outside the band of interest, so we regard the \citet{cohen04} measurement as an upper limit.  Moreover, if each pixel contains many unresolved sources, we would expect the net $\delta \zeta$ to be even smaller.  Thus the typical error in the 21 cm power spectrum resulting from our spectral fits is \citep{zald04}
\bq
\VEV{\delta T_{\rm fg}^2} \approx 0.04 \left( \frac{\bar{\delta T}_{\rm fg}}{40 \mkel} \, \frac{\delta \zeta}{0.2} \, \frac{|\nu_1 - \nu_2|}{6 \MHz} \, \, \frac{150 \MHz}{\nu_1} \right)^2 \mkel^2,
\label{eq:synch-error}
\eq
where $\bar{\delta T}_{\rm fg}$ is the mean brightness of all the foregrounds that survive the cross-correlation.  We see that the errors from foregrounds within the survey volume remain well below the signal for any reasonable frequency separation, so even the simplest foreground removal algorithm -- subtracting a constant from each $\bk_\perp$ pixel -- should suffice for measuring the 21 cm-galaxy cross-correlation.  This should be contrasted with the 21 cm measurement on its own, which requires more sophisticated algorithms because it must contend with contamination from sources at \emph{all} redshifts (e.g., \citealt{santos05, wang05, mcquinn05-param}).

\section{Discussion}
\label{disc}

We have studied the potential for measuring the cross-correlation between the redshifted 21 cm signal from the high-redshift IGM (during and before reionization) and the galaxy population.  We have emphasized four advantages to this measurement that will complement (from a data analysis perspective) observations of fluctuations in the 21 cm signal.  First, at least for the initial generation of experiments, the combination with a galaxy survey increases the expected signal-to-noise by a factor of a few.  This allows measurements to extend to somewhat smaller scales and significantly increases the dynamic range (in $k$-space) of 21 cm observatories.  Second, the galaxy survey modes have higher sensitivity along the plane of the sky and so improve measurements of the anisotropy of the 21 cm signal.  This will be useful in extracting fundamental cosmological parameters, because the redshift space distortions are sourced directly by the dark matter distribution and are much more robust to astrophysical uncertainties \citep{barkana05-vel, mcquinn05-param}.  

Third, by cross-correlating with galaxies (which we can unambiguously determine to be at the proper redshift), this measurement confirms the reality of the cosmological 21 cm signal.  Given the enormous brightness of the Galactic and extragalactic foregrounds (easily exceeding $\sim 100$--$1000 \kel$, compared to the $\sim 10 \mkel$ signal), and the complexity of the data analysis (see, e.g., \citealt{furl06-review}, \S 9), this conceptual advantage should not be underestimated.  Finally, measuring $P_{21,g}$ drastically reduces the requirements for foreground cleaning, because only those foregrounds originating from the survey volume (such as free-free and synchrotron emission from the galaxies; \citealt{oh03}) will survive the cross-correlation.  We have shown that, although modes in the plane of the sky will remain contaminated (albeit only at an order unity level), residual contamination of modes with a line-of-sight component will be small, even without any sophisticated cleaning algorithm.   

Although the ideal experiment would use a galaxy survey spanning the entire volume of the 21 cm observation, the signal-to-noise requirements are lax enough that even a much smaller galaxy survey can be interesting.  In fact, because 21 cm surveys are essentially thin redshift slices, modes wider than the redshift depth of the survey are lost anyway during foreground removal.  Thus, provided that the angular size of the galaxy survey exceeds the redshift depth, there is no disadvantage (other than a loss of statistical power) to using small fields -- the same range of $k$ modes is still available.  A survey of a few square degrees -- already achievable at $z \sim 6$--$7$ with present technology -- would satisfy this requirement.  For example, as shown by \citet{wyithe06-cross}, the existing Subaru Deep Field observations of $z=6.56$ \lya emitters \citep{kashikawa06}, in combination with an MWA field, could offer interesting constraints.

On the other hand, we have also found that, to take full advantage of the cross-correlation, spectroscopic redshifts are probably required.  This is because 21 cm arrays are primarily sensitive to line-of-sight modes, precisely those that redshift errors contaminate.  Thus the technical requirements for our surveys are substantially greater than, e.g., weak lensing, which only requires reasonably good photometric redshifts (so long as the distributions are understood extremely well).

Measuring the cross-correlation also drives the design of low-frequency telescopes in specific directions.  In particular, statistical measurements like the power spectrum have a tradeoff between survey area and depth in any particular field.  Either way is a viable method -- for example, the MWA (which has a large field of view) and LOFAR (which has much more collecting area, but a smaller field of view) offer comparable constraints on the 21 cm power spectrum \citep{mcquinn05-param}.  However, given the difficulty of deep, wide galaxy surveys, we are unlikely to be able to cover hundreds of square degrees in the near future.  The extra field of view of the radio telescope is then useless in regard to the cross-correlation:  for these purposes, depth is more important than field of view.  Another implication is that long baselines, even with a low filling factor, can be quite useful.  Because the galaxy power spectrum can extend to much smaller scales than the 21 cm survey's angular coverage, it can pull out extra information on the small-scale behavior -- but only if at least some baselines exist at the relevant scales.

In this paper, we have focused on the detectability of this cross-correlation rather than the physical insight to be gleaned from it.  Of course, the cross-correlation has obvious scientific uses, especially with regard to the relationship between the ionizing sources and the 21 cm line.  The most basic observable is whether over- or underdense gas is ionized first (``inside-out" versus ``outside-in" reionization), which should be measurable with the first surveys \citep{wyithe06-cross}.  However, there is of course much more to be gained:  for example, isolating those galaxies responsible for the bulk of the ionizing photons and testing the role of recombinations.  Another interesting application is the cross-correlation with different galaxy samples; for example, \citet{gunn65} absorption is expected to affect the \lya lines of galaxies during reionization \citep{miralda98-lya, haiman02-lya}.  But the amount of absorption suffered by each galaxy depends on how near it is to ionized gas -- so \lyans-selected galaxies can indirectly map the distribution of \hone and ionized bubbles in the IGM \citep{furl04-lya, furl06-lyagal}.  Cross-correlation of such a population with the 21 cm emission would help to isolate the \htwo region structure.  Given the complexity of the 21 cm signal (e.g., \citealt{furl04-bub, zahn06-comp}), these aspects are best explored with numerical simulations, which we defer to the future.

The cross-correlation will also, of course, be useful even before reionization gets underway.  In earlier phases, the galaxies seed fluctuations in the spin temperature (through \lya coupling; \citealt{barkana05-ts}) and the gas temperature (through X-rays; \citealt{pritchard06}), both of which affect the 21 cm brightness temperature.  Thus, eventually we hope to push both galaxy surveys and the 21 cm observations to much higher redshifts:  this will no doubt be difficult, given the rapid decline in the galaxy number density and the rapid increase in the 21 cm foregrounds.  But understanding the detailed properties of these earlier generations of galaxies would be an enormous payoff for the effort.

\acknowledgements

We thank M. McQuinn and O. Zahn for helpful discussions and comments on the manuscript.

\bibliographystyle{apj}
\bibliography{Ref_21cm,galcross}

\end{document}